\documentclass[12pt]{article}
\usepackage{epsf}
\title{ {\bf  Lepton Polarization Asymmetry in $B\rightarrow\ell^{+}\ell^{-}$ decay Beyond the Standard
Model}}

\author{\vspace{1cm}\\
        {\bf V. Bashiry}
         \thanks{E-mail address:
         bashiry@newton.physics.metu.edu.tr}
         \\
        Physics Department, Middle East Technical University \\
        Ankara, Turkey\\}
\date{}
\begin{document}
\setlength{\baselineskip}{24pt}
\maketitle
\setlength{\baselineskip}{7mm}
\begin{abstract}
 The lepton polarization asymmetry in the $B\rightarrow\ell^{+}\ell^{-}$ decay,
when one of the leptons is polarized, is investigated using the
most general form of the effective Hamiltonian. The sensitivity of
the asymmetry to the new Wilson coefficients is studied.Moreover,
correlations between the lepton polarization asymmetry and the
branching ratio is studied. It is observed that, there are not
exist such regions of new Wilson coefficients, which the value of
branching ratio coincides with SM result while the lepton
polarization does not, i.e new physics effects can be established
 by studying lepton polarization only.
\end{abstract}
\thispagestyle{empty}
\newpage
\setcounter{page}{1}
\section{Introduction}
The study of rare B-decays is one of the most important research
areas in particle physics. These decays induced by flavor changing
neutral currents (FCNC) and provide a promising ground for testing
the structure of weak interactions. These decays are forbidden in
the standard model(SM) at tree level and for this reason represent
" very good laboratory " for checking predictions of the SM at
loop level. Moreover,  these decays are very sensitive to the new
physics beyond the SM, since loop with new particles can give
considerable contribution to SM result. The new physics effects in
rare decays can appear in two different ways, namely modification
of Wilson coefficients existing is SM or through new operators
with new Wilson coefficients which absent in SM.\\
 The rare pure leptonic $B_{q}\rightarrow \ell^{+} \ell^{-} (q = d, s$ and $\ell
= e,\mu, \tau$ )decays  are very good probes to test new physics
beyond the standard model , mainly to reveal the Higgs
sector[1-3].\\ In aim of the present work is , investigation of
the lepton polarization as a tool for establishing new physics
beyond the SM, using the most general form of effective
Hamiltonian. More precisely our goal is following: Can we find
such a regions of new Wilson coefficients for which lepton
polarization differ from SM prediction, while
branching ratio  coincides with SM result.\\
Note that lepton polarization for $B_{q}\rightarrow \ell^{+}
\ell^{-}$ decay is studied in \cite{4}.\\
The paper is organized as follows: In Section 2 , we present the
theoretical expression for the decay widths and lepton
polarizations. Section 3 is devoted to numerical analysis and
conclusion.
\section{Double-Lepton polarization Asymmetry}
In this section we obtained the expression for decay width and
lepton polarization asymmetry, using the more general model
independent form of effective hamiltonian. The effective
Hamiltonian for the $b\rightarrow s \ell^{+}\ell^{-}$ transition
in terms of twelve model independent four Fermi interactions can
be written in following form \cite{5,6}
\begin{eqnarray}
{\cal{H}}_{eff}&=&\frac{G_{F}\alpha}{\sqrt{2}\pi}V_{ts}V^{*}_{tb}
\Bigg\{C_{SL}\overline{s}i\sigma_{\mu\nu}\frac{q^{\nu}}{q^{2}}\,L\,
b\overline{\ell}\gamma^{\mu}\ell\,+
C_{BR}\overline{s}i\sigma_{\mu\nu}\frac{q^{\nu}}{q^{2}}\,R\,
b\overline{\ell}\gamma^{\mu}\ell \nonumber \\ &+&
C^{tot}_{LL}\overline{s}_{L}\gamma_{\mu}b_{L}\overline{\ell}_{L}\gamma^{\mu}\ell_{L}+
C^{tot}_{LR}\overline{s}_{L}\gamma_{\mu}b_{L}\overline{\ell}_{R}\gamma^{\mu}\ell_{R}+
C_{RL}\overline{s}_{R}\gamma_{\mu}b_{R}\overline{\ell}_{L}\gamma^{\mu}\ell_{L}
\nonumber \\ &+&
C_{RR}\overline{s}_{R}\gamma_{\mu}b_{R}\overline{\ell}_{R}\gamma^{\mu}\ell_{R}+
C_{LRLR}\overline{s}_{L}b_{R}\overline{\ell}_{L}\ell_{R}+
C_{RLLR}\overline{s}_{R}b_{L}\overline{\ell}_{L}\ell_{R}
\nonumber\\ &+&
C_{LRRL}\overline{s}_{L}b_{R}\overline{\ell}_{R}\ell_{L}+
C_{RLRL}\overline{s}_{R}b_{L}\overline{\ell}_{R}\ell_{L}+
C_{T}\overline{s}\sigma_{\mu\nu}b\overline{\ell}\ell
\nonumber\\&+&
iC_{TE}\epsilon^{\mu\nu\alpha\beta}\overline{s}\sigma_{\alpha\beta}b\overline{\ell}\ell\Bigg\}
\,\,\, \label{Hamiltonian}
\end{eqnarray}
Where L and R in (1) are
\begin{eqnarray}
R=\frac{1+\gamma_{5}}{2}, \,\,\,\,\,\,\,\,\, L
=\frac{1-\gamma_{5}}{2} \nonumber
 \, , \label{LR}
\end{eqnarray}
 and $C_{x}$ are the coefficients of the four–Fermi interactions and $q = p_{2}+p_{1}$ is the momentum
transfer. Among twelve Wilson coefficients some of them are
already exist in the SM. For example, the coefficients $C_{SL}$
and $C_{BR}$ in penguin operators correspond to
$-2m_{s}C^{eff}_{7}$ and $-2m_{b}C^{eff} _{7}$ in the SM,
respectively. The next four terms in Eq. (1) are the vector type
interactions with coefficients $C^{tot} _{LL}, C^{tot} _{LR},
C_{RL}$ and $C_{RR}$. Two of these vector interactions containing
$C^{tot} _{LL} $and $C^{tot} _{LR}$ do exist in the SM as well in
the form $(C^{eff} _{9} - C_{10})$ and $(C^{eff} _{9} + C_{10})$.
Therefore we can say that $C^{tot} _{LL} $and $C^{tot} _{LR}$
describe the sum of the contributions from SM and the new physics
and they can be written as
\begin{eqnarray}
C^{tot} _{LL} = C^{eff} _{9} - C_{10} + C_{LL}\nonumber\, ,\\
C^{tot} _{LR} = C^{eff} _{9} + C_{10} + C_{LR}\nonumber\, ,
\end{eqnarray}\label{ctot}
The terms with coefficients $C_{LRLR}, C_{RLLR}, C_{LRRL}$ and$
C_{RLRL}$ describe the scalar type interactions. The last two
terms with the coefficients $C_{T}$ and $C_{TE}$, obviously,
describe the tensor type interactions. The amplitude of exclusive
$B\rightarrow \ell^{+} \ell^{-}$ decay is obtained by sandwiching
 of effective Hamiltonian between  meson and vacuum
states. It follows from Eq. (\ref{Hamiltonian}) that in order to
calculate the amplitude of the
 $B\rightarrow \ell^{+} \ell^{-}$ decay,
  following matrix elements are
 needed:
\begin{eqnarray}
\langle0|\overline{s}\,\gamma_{\mu}\gamma_{5}\,b|B\rangle&=&-i
f_{Bs}p_{\mu} \nonumber
\, , \\
 \langle0|\overline{s}\,\gamma_{5}\,b|B\rangle&=& i
f_{Bs}\frac{m_{Bs}^{2}}{m_{b}+m_{s}}\, ,\label{matrix}
\end{eqnarray}
All remaining matrix elements $\langle0|\overline{s}
\,\Gamma_{i}\, b|B\rangle$, where is one of the Dirac matrices $
I\
,\gamma_{\mu}\ , \sigma_{\alpha\beta}$ are equal zero. \\
For the matrix element of $B\rightarrow \ell^{+} \ell^{-}$ decay
we get
\begin{eqnarray}
M = i
f_{B}\frac{G_{F}\alpha}{2\sqrt{2}\pi}V_{ts}V^{*}_{tb}\Bigg[C_{PV}\,
\overline{\ell}\,\gamma^{5}\ell\,+\,
C_{PS}\,\overline{\ell}\,\ell\Bigg]\label{transamp}
\end{eqnarray}
 where
 pseudovector coefficient $C_{PV}$ and pseudoscalar coefficient
$C_{PS}$ are as following:
\begin{eqnarray}
C_{PV}&=& m_{\ell}(C_{LL}^{tot}-C_{LR}^{tot}-C_{RL}+C_{RR})+
\frac{m^{2}_{B}}{2(m_{b}+m_{s})}(C_{LRLR}-C_{RLLR}-C_{LRRL}+C_{RLRL})
\nonumber\, ,\\
C_{PS}&=&\frac{m^{2}_{B}}{2(m_{b}+m_{s})}(C_{LRLR}-C_{RLLR}+C_{LRRL}-C_{RLRL})\,,
\label{coef}
\end{eqnarray}
After some calculation we get following expression for the un
polarized $B\rightarrow \ell^{+} \ell^{-}$ decay width
\begin{eqnarray}
\Gamma_{0}&=&f^{2}_{B}\frac{1}{16\ \pi\ m_{B}}\Bigg|
\frac{G_{F}\alpha} {2\sqrt{2}\pi}V_{tb}\
V^{*}_{ts}\Bigg|^{2}\Bigg\{2\ C^{2}_{PV}\ m^{2}_{B} + 2\
C^{2}_{PS}\ m^{2}_{B}\ \upsilon^{2}\Bigg\}\upsilon
    \label{Gamazero}
\end{eqnarray}
where $\upsilon=\sqrt{1\ -\ m^{2}_{\ell}/m^{2}_{B}} $ is the final
lepton velocity.\\
Now let get expression for the lepton polarization. In the rest
frame of final leptons one can define only one direction.
Therefore the unit vectors of each lepton polarization can defined
as
\begin{eqnarray}
s^{\mu}=(0,\ \overrightarrow{e}_{L}^{\mp})=(0,\
\mp\frac{\overrightarrow{p}_{-}}{|\overrightarrow{p}_{-}|})
 \label{smu}
\end{eqnarray}
 where  is the tree momentum of $\ell^{-}$ and subscript L means longitudinal polarization.
 Boosting these unit vectors to the dilepton center of mass frame
 by using Lorentz transformation we get
\begin{eqnarray}
s_{\ell^{\mp}}^{\mu}=(\frac{|\overrightarrow{p}_{-}|}{m_{\ell}},\
\mp\frac{E_{\ell}\overrightarrow{p}_{-}}{m_{\ell}|\overrightarrow{p}_{-}|})
\label{smuboost}\\
\end{eqnarray}
where $E_{\ell}$ is the lepton energy.\\
 The decay width of the $B\rightarrow \ell^{+} \ell^{-}$ decay can
 written in following form
\begin{eqnarray}
\Gamma\ =\ \frac{1}{2}\ \Gamma_{0}\ \{1\ +\ P^{\mp}_{L}\
\overrightarrow{e}_{L}^{\mp}\ . \overrightarrow{n}^{\mp}\}
\label{gamapol}\\
\end{eqnarray}
where $P_{L}$ is longitudinal lepton polarization asymmetry. It
define as follows:
\begin{eqnarray}
P^{\mp}_{L}\ =\ \frac{\Gamma(\overrightarrow{n}^{\mp}\ =\
\overrightarrow{e}_{L}^{\mp})\ -\ \Gamma(\overrightarrow{n}^{\mp}\
=\
-\overrightarrow{e}_{L}^{\mp})}{\Gamma(\overrightarrow{n}^{\mp}\
=\ \overrightarrow{e}_{L}^{\mp})\ +\
\Gamma(\overrightarrow{n}^{\mp}\ =\
-\overrightarrow{e}_{L}^{\mp})}
 \label{pl}
\end{eqnarray}
The explicit expression of longitudinal polarization asymmetry is:
\begin{eqnarray}
P^{\mp}_{L}\ =\ \frac{2\ Re(C_{PV}\ C^{*}_{PS})\
\upsilon}{C_{PS}^{2}\ \upsilon^{2}\ +\ C^{2}_{PV}}
\label{plexplicite}
\end{eqnarray}
From this expression it is obvious that in SM lepton polarization
asymmetry  $P^{\mp}_{L}\ =\ 0$ since in SM $C_{PS}\ =\ 0$ (see eq.
 (\ref{coef})).
%

\section{Numerical analysis}.
In this section, we study the dependency of $P_{L}$ on new Wilson
 coefficients. In the present work all new Wilson coefficients are taken to be
 real. Here we would like to made following remark. Recent experimental
 results on the $B$ meson decay into two pseudoscalar meson
 indicated that Wilson coefficient $C_{10}$ can has large
 phase[7]. Therefore in principal appear new source for $CP$
 violating effects. We will discuss this possibility elsewhere.
  In performing numerical analysis we will vary the new
 Wilson coefficients describing the scalar interactions, in the
 range $-4\leq |C_{ii}|\leq 4$. The experimental result on
 branching ratio of  $B\rightarrow K (k^{*})\ell^{+} \ell^{-}$
 \cite{8,9} and the bound on branching ratio of the  $B\rightarrow \ell^{+} \ell^{-}$
 \cite{10} decay suggest that this is the right order of magnitude
 for scalar interaction.\\
 Now we are ready to perform numerical calculations. The values of
 input parameters which we have used in our numerical analysis
 are: \\ $ f_{Bs}=0.245$ Gev  \cite{11}, $ m_{B}=5.279.2 \pm\ 1.8$
 Mev,
 $ m_{\mu}=105.7$ Gev,  $ m_{\tau}=1777 $ Mev, $ \alpha=\frac{1}{129}$ \\
 The values of these parameters taken from
 \cite{12}.\\
 In Fig. 1 we present the dependence of longitudinal polarization
 of the lepton on Wilson coefficients of scalar interactions $C_{LRLR},\,
 C_{RLLR},\, C_{LRRL}\,$ and $C_{RLRL}$ for $B\rightarrow \mu^{+}
 \mu^{-}$decay. It should be noted that zero value of Wilson
 coefficients for scalar interactions corresponds to the standard
 model, case. \\
 From this figure we see that contributions coming from $C_{RLRL}$
 and $C_{LRLR}$ also, $C_{LRRL}$ and $C_{RLLR}$ are equal in
 magnitude but differ with sign. The similar circumstance take
 place for $B\rightarrow \tau^{+}
 \tau^{-}$decay (see Fig. 2). Therefore measurement the magnitude  and
 sign of the lepton  polarization can give unambiguous information
 about nature of scalar interaction.\\
 Obviously, if new physics beyond the SM exist, their effects can be
 appears in branching ratio, besides the lepton polarization. It
 is well known that the measurement of the branching ratio is more
 easy, that the lepton polarization. For this reason, it is more
 convenient and easy to study to study the branching ratio than
 the polarization, for establishing new physics beyond the
 standard model.\\
 In this connection we could like to discuss following problem:
 Can be establish new physics only by measuring lepton
 polarization. In other world, do exist such a regions of new
 Wilson coefficients, for which branching ratio coincides with
 the SM prediction, while lepton polarization do not. In order to
 answer this question, we study the correlations of single lepton
  polarization and branching ratio (see Fig. 3 and Fig. 4).\\
  From Figs. 3 and 4 we see that there are not exist such for
   regions of Wilson coefficients for which
 branching ratios coincides with the SM result, while lepton
 polarization do not. \\
 In summary, we present analysis for the longitudinal lepton
 polarization using the most general form of the effective
 Hamiltonian. We found that measurement of lepton polarization
 can provide us essential information about nature of scalar
 interaction. Moreover, we obtained that there are not exist such regions for the new Wilson coefficients, for which
 the only measurement of the lepton polarization gives invaluable
 information in looking for new physics beyond the SM.

\section{Acknowledgement}
The author thanks  Prof. Dr. TM Aliev and  E.O. Iltan for helpful
discussions.

\newpage
{\bf Figure Captions}\\
{\bf Fig. (1)}The dependence of longitudinal polarization
asymmetry $P_{L}$ on new Wilson coefficients responsible for
scalar interactions for the $B\rightarrow \mu^{+} \mu^{-}$ decay.
Here the solid, dashed, doted and small dashed lines corresponds
to $C_{LRLR}$, $C_{RLLR} $,$C_{RLRL}$ and $C_{LRRL}$,
respectively.\\
{\bf Fig. (2)} The same as {\bf Fig. (1)}, but for the
$B\rightarrow \tau^{+} \tau^{-}$ decay.\\
{\bf Fig. (3)} Parametric plot of the correlation between
longitudinal lepton polarization asymmetry and branching ratio for
the $B\rightarrow \mu^{+} \mu^{-}$ decay. The vertical line
corresponds to SM result
 for branching ratio. In this figure, solid line corresponds to
 the Wilson coefficients $C_{LRLR}$ and $C_{RLRL}$ , and dashed
 line $C_{LRRL}$ and $C_{RLLR}$, respectively.\\
{\bf Fig. (4)} The same as {\bf Fig. (3)}, but for the
$B\rightarrow \tau^{+} \tau^{-}$ decay.\\
\newpage
\begin{figure}[1]
\vskip -3.0truein \centering \epsfxsize=6.8in
\leavevmode\epsffile{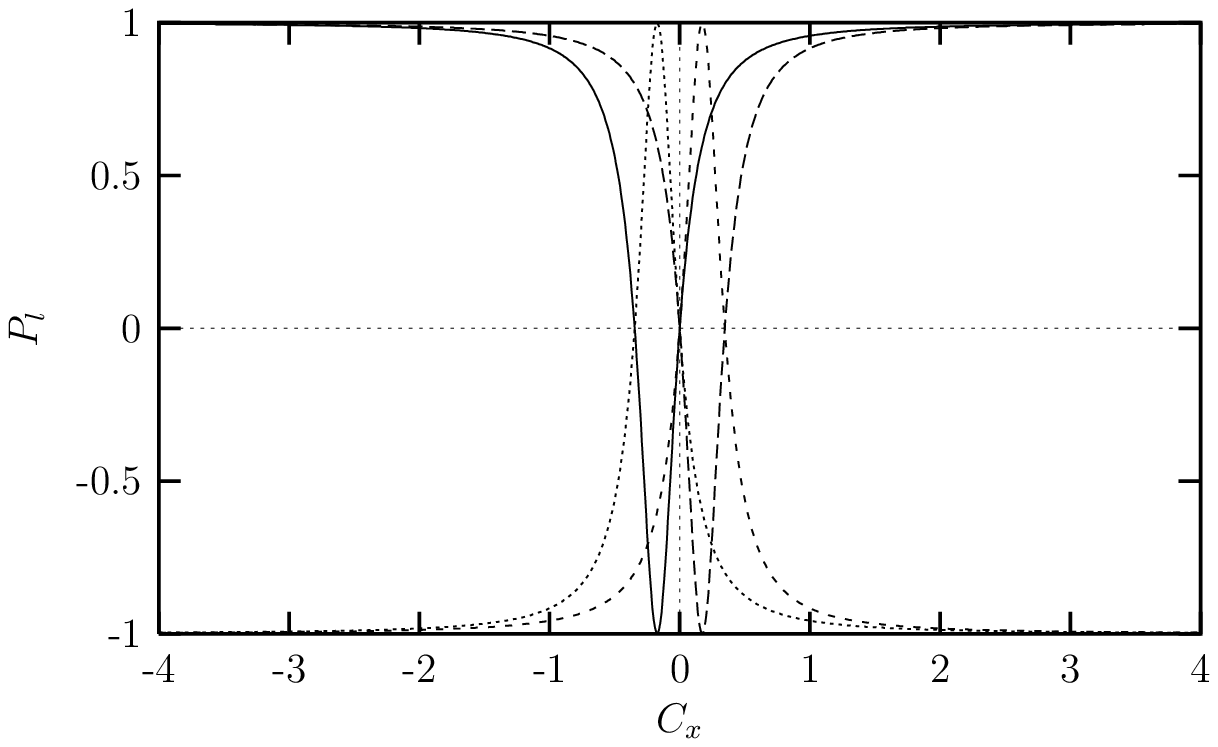} \vskip -4.0truein \caption[]{}
\label{plmion}
\end{figure}
\begin{figure}[2]
\vskip -3.0truein \centering \epsfxsize=6.8in
\leavevmode\epsffile{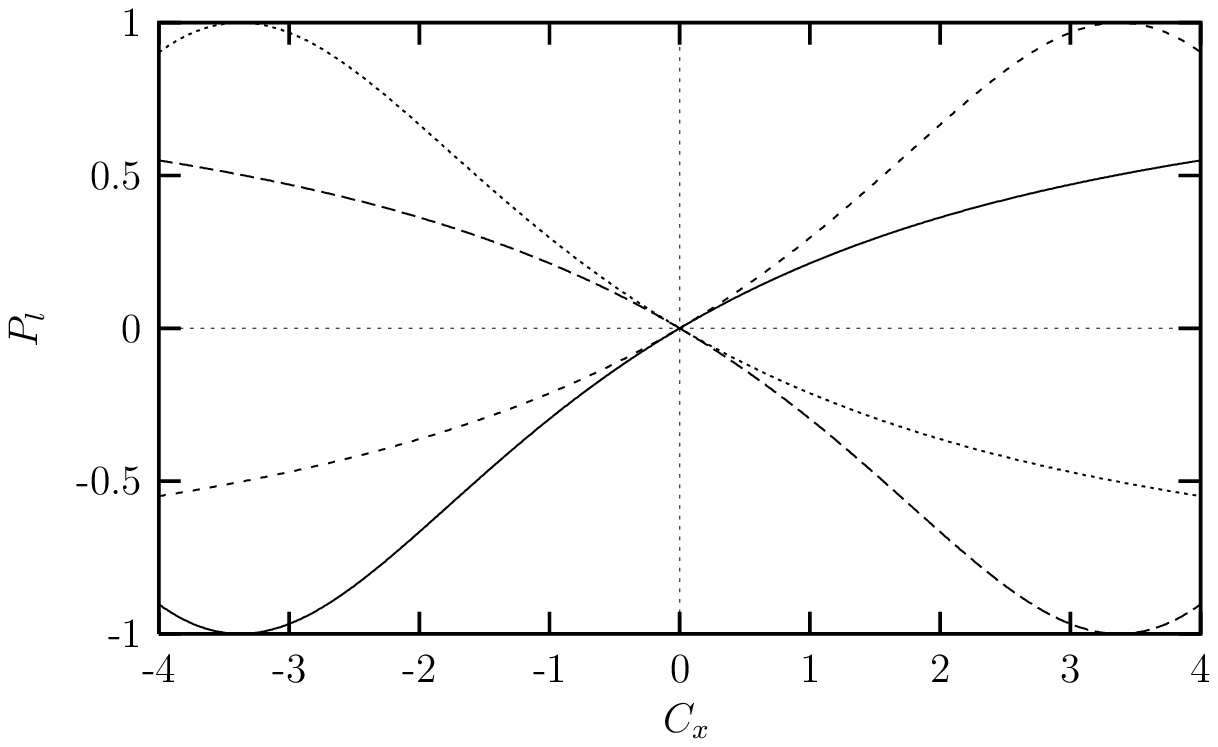} \vskip -4.0truein \caption[]{}
\label{pltau}
\end{figure}
\newpage
\begin{figure}[3]
\vskip -3.0truein \centering \epsfxsize=6.8in
\leavevmode\epsffile{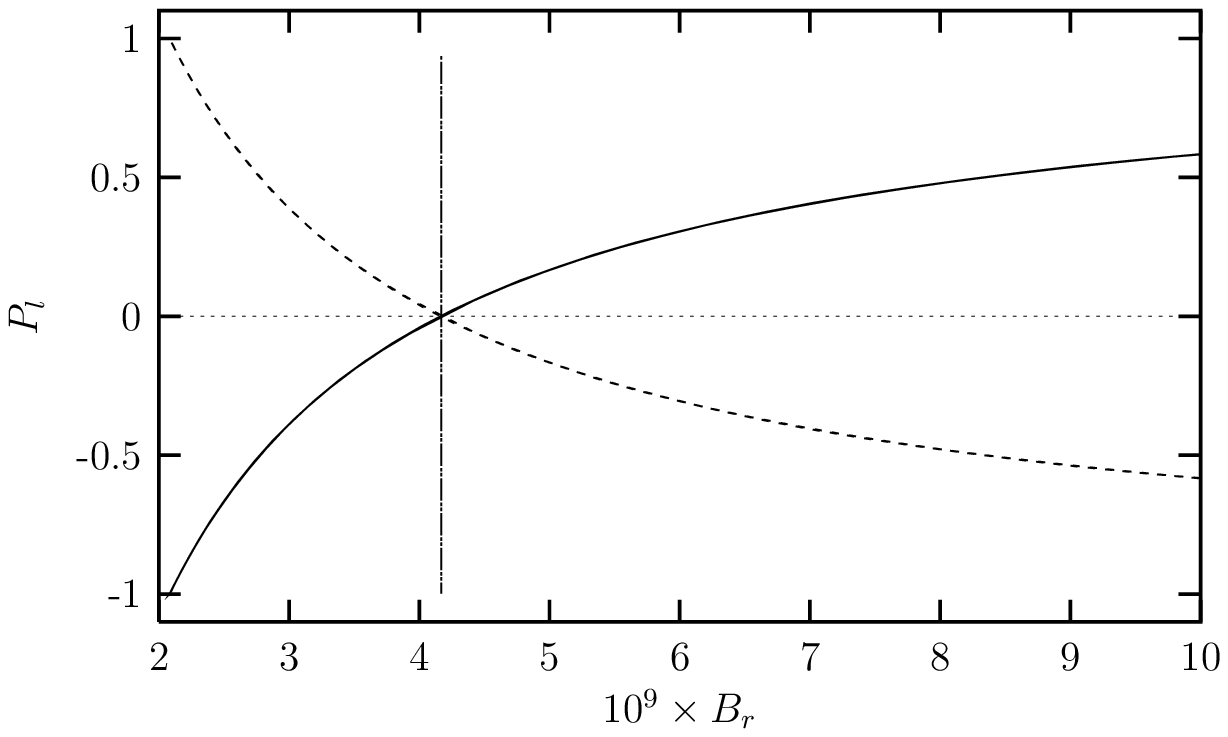} \vskip -4.0truein \caption[]{ }
\label{mion}
\end{figure}
\begin{figure}[4]
\vskip -3.0truein \centering \epsfxsize=6.8in
\leavevmode\epsffile{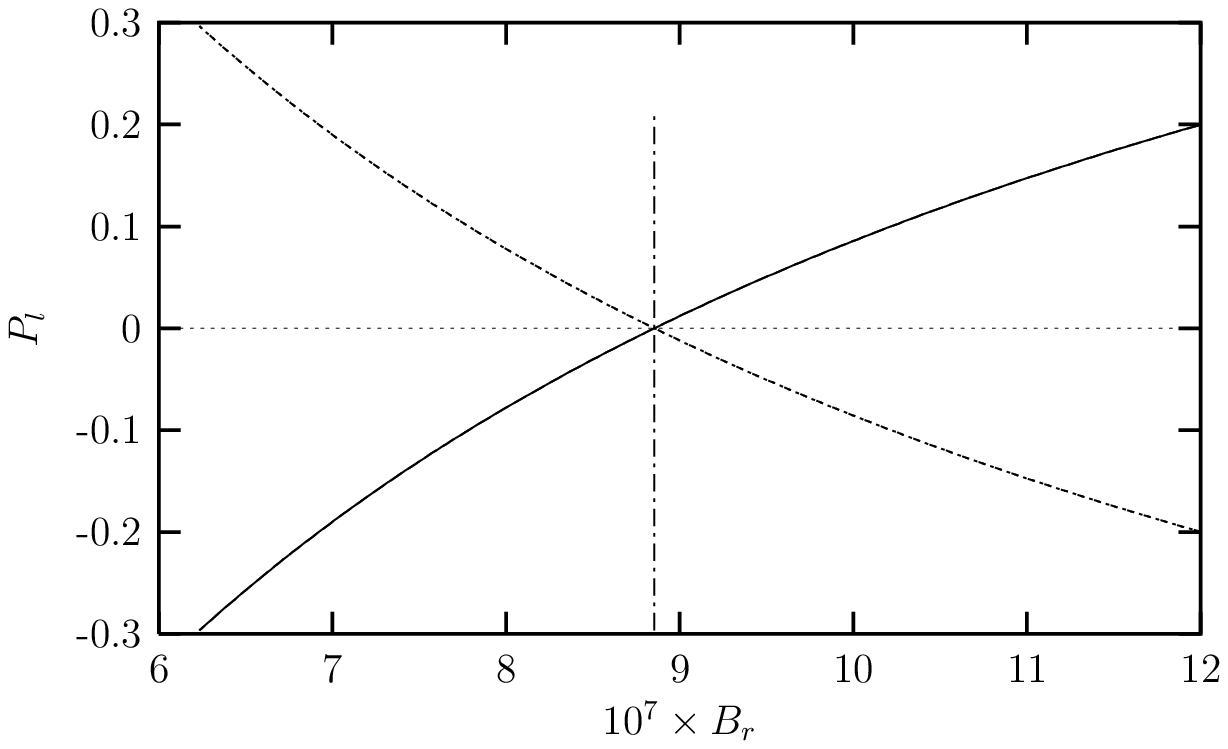} \vskip -4.0truein \caption[]{}
 \label{tau}
\end{figure}
\end{document}